\documentclass{PoS}

\title{INTEGRAL results on gamma-ray bursts and soft gamma-ray repeaters}

\ShortTitle{INTEGRAL results on gamma-ray bursts and soft gamma-ray repeaters}

\author{\speaker{Sandro Mereghetti}%
         \\
        INAF - IASF Milano, Italy\\
        E-mail: \email{sandro@iasf-milano.inaf.it}}

\abstract{Despite not being specifically designed for the study of gamma-ray bursts (GRBs),
the INTEGRAL satellite is giving a relevant contribution  to this field
with the detection of $\sim$10 GRB year$^{-1}$ in the IBIS field of view.
The ground-based INTEGRAL Burst Alert System (IBAS) has provided real-time localizations
with 2-3 arcmin accuracy for most of these events. The INTEGRAL sample now comprises 81 bursts,
including 4 of the short class.
IBAS has also revealed other  transient events from various classes of Galactic sources,
most notably  short bursts and flares from magnetar candidates.
Here I present a summary of the IBAS performances
in the eight years since the INTEGRAL launch and review some of the INTEGRAL results
for GRBs and Soft Gamma-ray Repeaters.
}

\FullConference{8th INTEGRAL Workshop ''The Restless Gamma-ray Universe''\\
		 September 27-30 2010\\
		 Dublin Castle, Dublin, Ireland}

\def\approxgt{\mathrel{\hbox{\rlap{\lower.55ex \hbox {$\sim$}}
        \kern-.3em \raise.4ex \hbox{$>$}}}}
\def\approxlt{\mathrel{\hbox{\rlap{\lower.55ex \hbox {$\sim$}}
        \kern-.3em \raise.4ex \hbox{$<$}}}}

\def\pdot {\dot P}

\def\ltsima{$\; \buildrel < \over \sim \;$}
\def\lsim{\lower.5ex\hbox{\ltsima}}
\def\gtsima{$\; \buildrel > \over \sim \;$}
\def\gsim{\lower.5ex\hbox{\gtsima}}

\def\sgr {1E\,1547.0--5408}
\def\src {SGR 1806--20}

\begin{document}

\section{Introduction}

INTEGRAL was successfully launched on 17 October 2002.
On November 25, less than one month after the switch-on of its main instruments,
INTEGRAL detected its first Gamma-ray Burst in the IBIS field of view (GRB 021125 \cite{malaguti2003}).
A few weeks later the first GRB localization in real time was obtained for GRB 021219 \cite{mereghetti2003}.
These results were not unexpected, since pre-launch
estimates \cite{mcb2001} predicted a rate of $\sim$1 GRB/month in the 0.23 sterad field of view of IBIS.

Up to now (January 2011), 81 GRBs  occurred in the IBIS field of view \footnote{See online list at
http://ibas.iasf-milano.inaf.it/IBAS\_Results.html}.
Most of them were discovered in near real time by the INTEGRAL Burst Alert Systems
(IBAS, \cite{ibas}), resulting in the prompt distribution of  their positions with accuracy
of $\sim$2-3 arcmin.
IBAS has also revealed transient events from other classes of sources, like type I X-ray bursts from
low mass X-ray binaries, short bursts and flares from magnetar candidates, and outbursts
from  other galactic transients.  Here I briefly review some of the INTEGRAL results
for GRBs and Soft Gamma-ray Repeaters localized with IBAS.

\section{Gamma-ray Bursts}

\subsection{IBAS performances}

\begin{figure}
\centering
\includegraphics[height=60mm,angle=0]{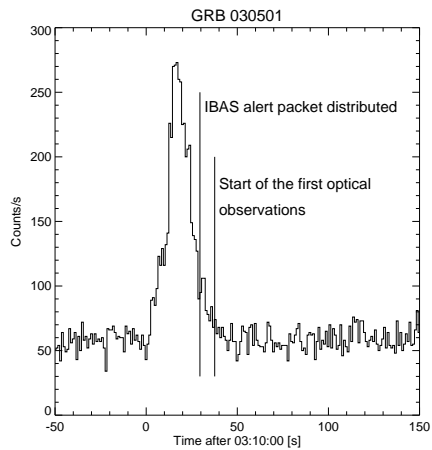}
\caption{GRB 030501, the first GRB with an arcmin localization distributed in real time.}
\label{fig-030501}
\end{figure}

The IBAS software for GRB detection runs in real time at the INTEGRAL Science Data Center
(ISDC \cite{cou03}), exploiting the fact that the satellite data are continuously downloaded and
received at the ISDC without significant delays.
IBAS started routine operations in April 2003, after a few months of testing and parameter tuning
during which three GRBs were detected, but the automatic delivery of their positions was not yet
enabled.
On May 1st, the coordinates of GRB 030501 \cite{bec03},
with an uncertainty of only 4 arcmin were distributed $\sim$30 s after
the start of  the burst  (see Fig. \ref{fig-030501}).
This was in absolute the first GRB with a small error region distributed in near real time
(Swift became fully operational in the first months  of 2005).

Rapid localizations have been obtained  for $\sim$60 INTEGRAL GRBs, while in  other $\sim$20
cases the coordinates have been distributed after some delay
because, to avoid generating too many false alerts, the threshold for automatic distribution
is kept at a conservative level. The triggers below such a threshold are checked
with an interactive analysis, and indeed for most of them it is not possible to distinguish
between a statistical fluctuation of the background and a genuine astrophysical event.
However, several real GRBs were found in these off-line analysis and their coordinates distributed with
typical delays of a few hours. Since February 2011 the sky coordinates also for low
significance IBAS triggers are distributed in real time to interested users.

The Optical Monitoring Camera (OMC)
on board INTEGRAL covers only the central 5$^{\circ}\times5^{\circ}$ around the
satellite pointing direction, providing images of predefined CCD windows
at the positions of known sources.
When IBAS finds that a GRB is in the OMC field of view, a telecommand is automatically uploaded
to the OMC in order to add a new CCD window covering  the burst position.
Up to now, this happened only for  GRB 050626,  which unfortunately
was at only  $\sim$3 arcmin from  $\alpha$ Crucis and the OMC
data were  completely saturated by this very bright star.

IBAS also reveals  GRBs in the light curves produced by the Anti-Coincidence Shield (ACS)
of the  SPI instrument, which, besides serving to reduce the background in the
germanium spectrometer, is routinely used as a nearly
omni-directional detector for GRBs \cite{von03}. It produces
light curves with a resolution of 50 ms for photon energies \gtsima 80 keV, but
without energy and directional information.
These lightcurves are used for GRB triangulations with other satellites.

\begin{table}[ht!]
\begin{center}
\begin{footnotesize}
\begin{tabular}{|l|c|c|}
\hline
GRB						& redshift	& reference \\
\hline
031203  				& 0.1055 & \cite{pro04} \\
050223  				& 0.584	 & \cite{pel06} \\
050502A  				& 3.793	 & \cite{pro05} \\
050525A  				& 0.606	 & \cite{fol05} \\
080603  				& 1.688	 & \cite{per08} \\
\hline
\end{tabular}
\end{footnotesize}
\caption{INTEGRAL GRBs with spectroscopic redshift}
\label{tab-z}
\end{center}
\end{table}

\begin{figure}
\centering
\includegraphics[height=70mm,angle=0]{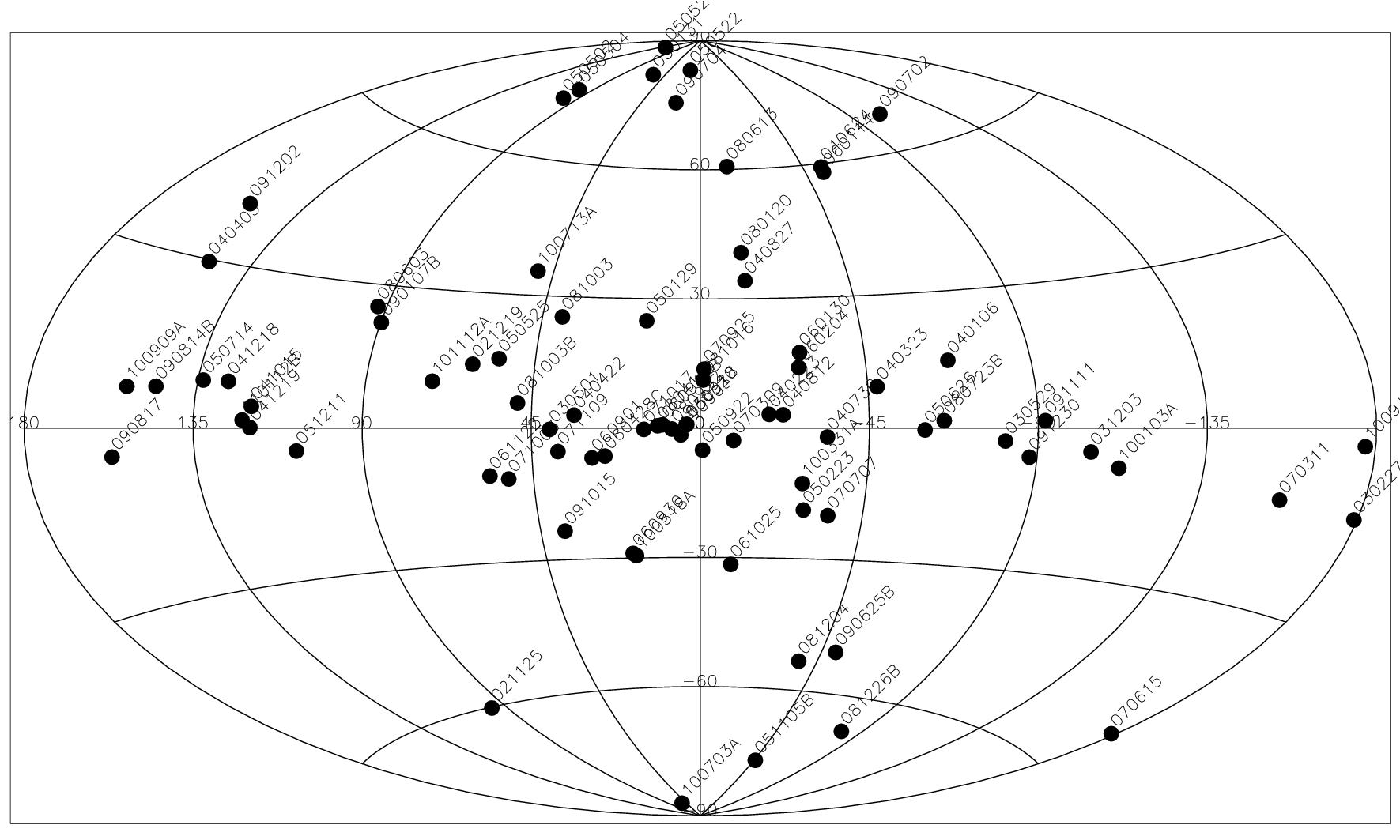}
\caption{Positions in galactic coordinates of the GRBs detected with INTEGRAL/IBIS.}
\label{fig-pos}
\end{figure}

\subsection{Global properties of the GRBs detected with IBIS}

The sky positions of the 81  GRBs located inside the IBIS  field of view are shown in galactic
coordinates in Fig.\ref{fig-pos}.
The anisotropy in their distribution is due to the non-uniform sky coverage
of the INTEGRAL pointing program, mostly devoted to the Galactic bulge and plane.
Only four of the INTEGRAL bursts belong to the class of short GRBs.
This is a much smaller fraction than in the BATSE sample, where short GRBs accounted for
about one fourth of the whole population,  and is more similar to the findings of other
instruments  with a greater sensitivity  at low energy than BATSE, like
Swift/BAT (short GRBs account for $\sim$10\% of the BAT sample).

Afterglows have been detected for more than half of the  INTEGRAL GRBs:
41  at X-ray energy, 21 in the optical, and 3 in the radio band.
Virtually all the GRBs for which follow-ups
were carried out with XMM-Newton or Swift had an X-ray afterglow detected.
Optical observations were in some cases discouraged by the low Galactic latitude.
This possibly explains the relative paucity of optical afterglows in the INTEGRAL sample,
but in some cases really deep upper limits were obtained (e.g. GRB 040223, GRB 040624 \cite{fil06},
and GRB 040403 \cite{mer05b}).

Spectroscopic redshifts have been obtained for 5 INTEGRAL GRBs (Table 1).
GRB 050502 is the INTEGRAL GRB with the highest redshift.
Its optical afterglow was detected only $\sim$23 seconds after the IBAS trigger
and follow-up spectroscopy with the Keck
telescope yielded a redshift of $z=3.793$ \cite{pro05}.
GRB 031203, at z=0.1, was one of the first GRBs with a spectroscopic supernova association, and
is discussed in more detail below.

A systematic analysis of all the bursts detected with IBIS before mid 2007  \cite{via09}
shows that, from the spectral point of view, they do not differ significantly from the larger
sample of Swift/BAT  GRBs.  In the relatively narrow energy
band covered by the ISGRI/IBIS detector, their spectra are generally well fit by a  power law with
average photon index $\sim$1.6.  This value is intermediate between the average low-energy and
high-energy slopes obtained with the broken power law model usually adopted for GRBs, suggesting
a spectral break within the IBIS/ISGRI range. Indeed this is confirmed for the few bright bursts
with sufficient statistics, for which an exponentially cut-off power-law gives a
significantly better fit and the average cut-off energy is $\sim$80 keV.
On the other hand, in a few cases the steep  power-law spectra measured by ISGRI (photon index $>$2.5)
indicate that the event can be classified as an X-ray Flash (e.g. GRB 030529 and 040903 \cite{via09}).

The peak fluxes of the   GRB revealed by IBIS range from  $\sim$0.08 ph cm$^{-2}$
s$^{-1}$ to $>$50 ph cm$^{-2}$s$^{-1}$  (20--200 keV, 1 s integration time).
The upper limit is uncertain because bright fluxes saturate the telemetry allocated to IBIS.
A possible difference between the INTEGRAL and Swift GRB samples is suggested by the comparison
of their brightness.
A probability of only 3\% that the IBIS and BAT peak fluxes are drawn from the same
distribution was found \cite{via09},  indicating that IBIS,  on average, reveals fainter GRBs than BAT.
This is consistent with the smaller field of view of IBIS, giving a higher sensitivity near the
on-axis direction, at the expenses of a lower rate of detections.

The spectral lag for a sample of INTEGRAL GRBs  was computed by \cite{fol08}, who found
that the subclass of faint bursts with long spectral lag are spatially correlated with the Supergalactic plane. This has been interpreted as evidence for a local population of low luminosity GRBs,
a result questioned by \cite{xia09} on the basis of the Swift sample.
Although a reanalysis of the INTEGRAL data, accounting for
the IBAS selection effects that were neglected in \cite{fol08},
supports the presence of this spatial correlation \cite{via09},
a larger sample is needed to unambiguously confirm the existence of such a local population.

\subsection{GRB 031203}

GRB 031203 is among the most interesting GRBs discovered by INTEGRAL.
Its afterglow  was observed at  X-ray, IR/Optical,
and radio wavelengths, and spectroscopic evidence of an associated Type Ic Supernova was found
\cite{mal04}.
Its 20-200 keV fluence of $\sim$10$^{-6}$ erg cm$^{-2}$  and  low  redshift (z=0.1)  imply an
isotropic-equivalent  energy E$_{iso}$ of only a few 10$^{49}$ erg,
three orders of magnitude below that of normal GRBs \cite{saz04}, and more
typical of an X-ray flash.
On the other hand, as confirmed in our re-analysis of the IBIS data with  recent
software and calibrations \cite{via09}, GRB 0312013 had a hard spectrum well fit
by a power-law with photon index $\sim$1.5, and a  99\% c.l. lower limit of 100 keV was derived  on its E$_{peak}$.
These results  make GRB 031203 an outlier of the correlations followed by most GRBs \cite{ama02,ghi04}.

The X--ray images obtained with \textit{XMM-Newton} a few hours after the burst led to the
discovery of  expanding  rings due to the scattering of the GRB
X-ray emission by dust grains in our Galaxy \cite{vau04}.
This  ``echo'', the first one observed around a GRB,  provides an indirect way to
estimate the intensity of the prompt  X--ray emission, which is found to exceed
the extrapolation to a few keV of the INTEGRAL spectrum \cite{vau04,tie06}.
Thus it is likely that the soft X--ray emission in GRB 031203 consisted of a
delayed component following the hard pulse seen with INTEGRAL,
similar to what has been observed in GRB 060218 \cite{ghi06}. In
this case GRB 031203 would have a smaller E$_{peak}$ and a higher
luminosity, making it in agreement with the ``standard'' relations.

\subsection{GRB  041219A}

GRB 041219A, the longest and one of the brightest GRB  localized by IBAS,
was composed of one weak precursor followed,  after a quiescent interval lasting $\sim$4 minutes,
by  the main GRB emission consisting of two bright pulses.
Thanks to the IBAS trigger and localization on the faint precursor,
successful optical and near IR observations started while the burst was still on going
\cite{2005Natur.435..181B,2005Natur.435..178V}.
The prompt optical  and gamma-ray emission were correlated,
suggesting that both components originated in the internal shock,
contrary to the only previously observed case, GRB 990123, in which the optical
prompt emission was  anticorrelated with the high-energy light curve, and was interpreted
as arising in the reverse shock.

The large fluence of GRB 041219A prompted searches for polarization exploiting Compton interactions
in the INTEGRAL instruments.
Hints for the presence of linear  polarization above 100 keV during the first and brightest pulse were
obtained from the
SPI data \cite{2007A&A...466..895M}, but with a low statistical significance.
These findings were only partially confirmed with  IBIS \cite{goe10},
which, on the other hand, showed  a possible linear polarization of the second pulse.
The analysis of IBIS data seems also to indicate that the polarization varies, both in intensity and angle,
during the burst.
Although these results are tantalizing and have potentially very interesting implications for
GRB theories, they are based on complicated and non-standard analysis, and their real significance,
which also should properly take into account the number of trials,  is not easy to assess.

\subsection{GRB 070707}

GRB 070707 was the first short GRB detected by INTEGRAL.
It had  a duration of 0.8 s, a spectrum well described by a power law with photon index
1.2, and a small spectral lag of 20$\pm$5 ms \cite{mcg08}.
These characteristics are typical of the class of short-hard bursts.
An X-ray afterglow was discovered with Swift,  permitting the identification of a
faint (R$\sim$23) optical transient.
The optical flux initially decayed very steeply, and then leveled at R$\sim$27 after a few days.
If, as it is likely, the constant flux is due to the host galaxy, this is one of the
faintest revealed for a short GRB \cite{pir08}.

\section{Soft Gamma-ray Repeaters}

Soft gamma-ray repeaters (SGRs) are X-ray sources characterized by the  emission,
during sporadic periods of activity, of short ($<$1 s)
bursts of soft gamma-rays with peak luminosity up to 10$^{42}$ erg s$^{-1}$, and,
much more rarely, by the occurrence of giant flares, releasing up to 10$^{46}$ ergs.
SGRs are well explained as magnetars: isolated neutron stars powered by extremely high magnetic fields,
$B>10^{14}$--10$^{15}$ G \cite{tho95}.
Short bursts have been detected also from most  Anomalous X-ray Pulsars
(AXPs, \cite{mer95}), thus supporting their interpretation as another class of magnetars. 
The most recent observations of AXPs and SGRs suggest that their different classification
merely reflects the way they were originally discovered (see \cite{mer08} for a review).

\subsection{SGR 1806--20}

\src\ is located at only 10$^{\circ}$ from the Galactic Center direction,
a sky region extensively observed by INTEGRAL, and
has been one of the most active SGRs after the INTEGRAL launch.
More than 300 bursts  have been detected, of which about 100 occurred within 10
minutes of exceptionally high activity on October 5, 2004 \cite{goe06}.
The high sensitivity of IBIS allowed us to  detect  the
faintest bursts ever observed from this SGR in the 20-100 keV energy
range, reaching fluences below S$\sim$10$^{-8}$  erg cm$^{-2}$.
The LogN-LogS of the bursts  is consistent with a single power law with slope 0.91$\pm$0.09,
independent of the bursting rate.
Time resolved spectroscopy of some of the brightest bursts showed  significant spectral
evolution, generally with the softest emission at the peak \cite{goe04}.

The giant flare emitted from \src\ on 2004 December 27 was the brightest
event of this class ever observed, and it was first reported thanks to its detection
with IBAS in the SPI/ACS data \cite{bor04}.
While all the other  satellites could observe the giant flare only for a few minutes before it
faded below their sensitivity, the  ACS detected high-energy
emission ($>$80 keV) from \sgr\ lasting about one hour after the giant flare \cite{mer05}.
This component, observed for the first time in a magnetar giant flare,
has been interpreted as a hard X--ray afterglow produced by the matter ejected relativistically,
with bulk Lorentz factor $\Gamma\sim$15 during the initial spike at the beginning of the flare.

\subsection{1E 1547--5408}

Although known since more than 30 years, the transient X-ray source \sgr\ was recognized as a member
of the magnetar class only recently, after the discovery of radio and X-ray pulsations
with a period of 2.1 s and period derivative $\pdot=2.3\times10^{-11}$ s s$^{-1}$,
and the observation of SGR like bursts \cite{gel07,cam07c}.

In January 2009 \sgr\ entered a new phase of bursting activity,
during which many short bursts of soft gamma-rays were emitted.
The period with the highest bursting rate was observed
with the ACS, which provided a continuous, uninterrupted
coverage from 14:35 UT of January 20  to 04:23 UT of January 23.
Two of the bursts showed tails lasting several seconds and modulated at the spin period
of \sgr\ resembling the pulsating tails seen after giant flares \cite{mer09,sav10}.
However, the energetics of these two bursts ($\sim$10$^{43}$ erg  for d=5 kpc) was not as large
as that of giant flares.

The peak of the bursting rate occurred around 6:48 UT of January 22,
when more than 50 bursts were recorded in 10 minutes.
The soft X-rays emitted during this  short time interval were scattered by three clouds of
interstellar dust, producing expanding halos around the source that were imaged with Swift/XRT and still detectable with XMM-Newton after twelve days \cite{tie10}.
Since the time evolution  of the halo flux and angular size
depend on the distances of the source and of the scattering clouds, it was possible to
derive the distance of \sgr\  and to get some
constraints on the dust properties and distribution from an analysis of the X-ray data.
Although the source distance depends on the adopted dust model, the best fits favored values of
$\sim$4-5 kpc, supporting the association of \sgr\ with the supernova remnant  G\,327.24--0.13.

\section{Conclusions}

Although INTEGRAL was designed as a multi-purpose gamma-ray Observatory, not specifically
optimized for GRBs, it has significantly contributed to this field thanks to its excellent
imaging performances and good sensitivity.
INTEGRAL unique features include the capability of long uninterrupted
monitoring with the SPI/ACS and the possibility to carry out polarization
searches exploiting Compton interactions in the multiple detectors.
Unfortunately, some of the potentials for INTEGRAL GRBs studies are hampered by limitations in the
satellite telemetry, which reduce the available count statistics for the brightest events.
The recent implementation of IBAS alert distribution for low significance triggers should lead
to a higher rate of INTEGRAL GRBs, possibly confirming the presence of a local population
of GRBs and/or  leading to the discovery of more bursts at high redshift.

\bigskip

\textbf{Acknowledgments.} I acknowledge all the scientists and engineers who
participated to the IBAS design, in particular
Jurek Borkowski and Diego G\"{o}tz for their invaluable contribution to its realization and operation.
Smooth running of IBAS has been  possible thanks to
M. Beck, V. Beckmann, N. Mowlavi, C. Ferrigno, E. Bozzo and all the operational staff of the ISDC.

\end{document}